\documentclass{article}
\usepackage[margin=1in]{geometry}
\usepackage[utf8]{inputenc}
\usepackage[numbers,sort&compress]{natbib}

\renewcommand{\bibsection}{\subsubsection*{\bibname}}

\usepackage[dvipsnames,table,xcdraw]{xcolor}
\usepackage{comment}
\usepackage{pdfcomment}
\usepackage{environ}
\RenewEnviron{comment}{\pdfcomment{\BODY}}
\usepackage{dcolumn}
\usepackage{booktabs}
\usepackage{tikz}
\usetikzlibrary{positioning,shapes,arrows}
\usetikzlibrary{decorations.pathreplacing}
\usepackage{graphicx}
\usepackage{subcaption}
\usepackage{algorithm}
\usepackage[noend]{algpseudocode}
\usepackage{amsmath}
\usepackage{amssymb}

\algrenewcommand\algorithmicindent{1em}

\usepackage{amsthm}
\theoremstyle{definition}

\theoremstyle{remark}

\makeatletter
\def\thm@space@setup{%
  \thm@preskip=0mm plus 0.5mm minus 0mm
  \thm@postskip=0mm plus 0.5mm minus 0mm 
}
\makeatother
\usepackage{enumerate}
\newcommand{\norm}[1]{\left\lVert#1\right\rVert}
\usepackage{wrapfig}
\usepackage{amssymb}

\title{Wireless localization with diffusion maps}
\author{Amin Ghafourian$^{\dag\ast}$, Orestis Georgiou$^\S$, Edmund Barter$^\ddag$, and Thilo Gross$^\dag$}
\date{\today}
\begin{document}
\maketitle
\vspace{-0.5cm}
\begin{center}
    \footnotesize
    $^\dag$ University of California, Davis, USA, $\S$ University of Cyprus, Nicosia, Cyprus, $^\ddag$ University of Bristol, UK \\
    $\ast$ corresponding author: aghafourian@ucdavis.edu 
 \end{center}

\begin{abstract}
In the Wireless Localization Matching Problem (WLMP) the challenge is to match pieces of equipment with a set of candidate locations based on wireless signal measurements taken by the pieces of equipment. This challenge is complicated by the noise that is inherent in wireless signal measurements. Here we propose the use of diffusion maps, a manifold learning technique, to obtain an embedding of positions and equipment coordinates in a space that enables coordinate comparison and reliable evaluation of assignment quality at very low computational cost. We show that the mapping is robust to noise and using diffusion maps allows for accurate matching in a realistic setting. This suggests that the diffusion-map-based approach could significantly increase the accuracy of wireless localization in applications.  
\end{abstract}

\section{Introduction}
The Internet of Things (IoT) is a developing technology that envisions creating interconnected systems whose components communicate and interact with each other and can be remotely monitored and controlled. These systems would blend seamlessly with the environment and could be incorporated in various industrial, domestic, military, and environmental applications. In particular, IoT systems are essential to the Smart City vision, where billions or even trillions of devices and sensors will be interconnected and form an enormous network \cite{gubbi2013internet,bryzek2013roadmap}. Case studies have demonstrated the positive impact of such technologies on society in domains such as transportation, infrastructure services, and public safety \cite{sanseverino2017smart}.

A particular class of interconnected systems are wireless sensor networks (WSN), i.e.~networks of low-power wireless devices (sensor nodes) that carry out measurements of the physical attributes of equipment or the environment \cite{nguyen2017wireless,park2019animal}. Each node is typically comprised of a microcontroller, memory, battery, a wireless RF transceiver and sensors to measure quantities such as pressure, humidity, and temperature. 

In applications, it is often important to keep a record of sensor nodes locations so that the received data can be accurately interpreted \cite{nguyen2017wireless}. While GPS modules can be used for this purpose, this incurs additional production costs and results in additional power consumption of the sensors (about 90mW). Furthermore, these modules do not perform sufficiently accurately indoors and underground, where there is poor satellite reception \cite{nguyen2017wireless}.

The limitations of using GPS modules define a need for alternative localization methods. RF-based methods broadly fall into the categories of database, range-based, range-free, and angle-based methods \cite{mao2007wireless,xiao2016survey,rai2012zee,savvides2001dynamic,niculescu2003dv,nguyen2015maximum,kotaru2015spotfi}. Database methods compare signal features, for instance Received Signal Strength Indicator (RSSI), to a pre-generated database and try to identify the most likely location of target devices \cite{rai2012zee}. Range-based methods estimate the distance of the target device from a number of known devices, or anchors, and multilaterate the location of the device \cite{savvides2001dynamic}. Range-free methods utilize RF multihop statistics of the network topology to determine the location of target devices \cite{niculescu2003dv,nguyen2015maximum}. In angle-based methods, multiple antennas are employed to identify the angle of arrival (AoA) of the RF signal and triangulate the location of target devices \cite{kotaru2015spotfi}. All of these methods are subject to errors due to background noise, wireless multipath fading, shadowing, non line-of-sight (NLoS), pathloss, etc. As a result, empirical models are used to retune and improve the accuracy of these methods \cite{nguyen2017wireless}.

An interesting idea was put forward by Keller and Gur \cite{keller2011diffusion} who propose to use a spectral embedding method for localization. This method provides a very elegant approach to the construction of a spatial map from pairwise measurements between nodes. However due to inherent difficulty of the challenge the locations of several nodes need to be known to achieve good results.

A particular incarnation of the wireless localization problem is the so-called Wireless Localization Matching Problem (WLMP) \cite{nguyen2017wireless2} where the challenge is to match a set of sensor nodes with an \textit{a priori} known set of sensor locations, called positions. This type of problem arises, for instance, in facilities where a large number of devices need to be installed in specific locations (e.g.~smart lights, smoke detectors, thermostats, etc.). In this setting the set of positions is known and documented in the facility blueprint, whereas keeping track of the position at which a particular device is installed (likely by subcontractors) creates a significant management overhead \cite{nguyen2017wireless,xiao2016survey}. A similar challenge arises in novel approaches to disaster response, where sensor nodes are airdropped, leading to a \textit{de facto} random distribution of nodes \cite{mao2007wireless,manshahia2016wireless}. The set of positions of nodes, but not individual node IDs, can then be obtained by aerial imaging such that locating each device again poses a WLMP \cite{nguyen2017wireless2}.

In this paper, we formulate a solution strategy for WLMP, which is in essence a bipartite matching problem between a number of positions and sensor nodes. In our proposed method, we use RSSI between pairs of nodes to obtain the best matching between nodes and position candidates. The key innovation that makes this matching efficient is the use of diffusion maps to embed the nodal positions in a new Euclidean space where matching is easy. In contrast to \cite{keller2011diffusion} the application of the diffusion map to the conceptually simpler WLMP challenge allows us to capitalize strongly on the power of the diffusion map and achieve very accurate results. The proposed solution is versatile and computationally efficient, which makes it an attractive approach for a spectrum of applications.

\section{Methods}
Matching problems are well studied and powerful algorithms that solve such problems exist. However, before these algorithms can be applied we must formulate a method to quantify how well a set of RSSI measurements match a specific position. A complicating factor is that measurements and positions use different coordinate systems. The RSSI measurements encode sets of pairwise distances, whereas the positions are given directly in terms of physical coordinates, e.g.~longitude and latitude. The positions thus live in a low-dimensional (typically 2D) physical space whereas the RSSI measurements 
live in a high-dimensional data space. This means that the distance estimates from RSSI, the positions, or both need to be mapped onto a different coordinate system before matching algorithms can be applied. The question then arises what coordinate system is most suited to facilitate the subsequent matching. 

The key insight used in this paper is that an advantageous coordinate system for the matching problem can be constructed using a method known as the diffusion map \cite{coifman2006diffusion,coifman2005geometric,lafon2004diffusion,lafon2006data,barter2019manifold}. The diffusion map is a manifold learning technique that can discover low-dimensional manifolds in datasets. Node positions in space can be thought of as one such dataset in which a one, two, or three-dimensional manifold exists. This is due to the expectation that for the purpose of the localization, node positions can adequately be described in a Euclidean space with as many dimensions. The diffusion map can then be used to find a natural parameterization of this manifold implied by the distances. The matching is then carried out in the new coordinates determined by the diffusion map (explained below). Importantly, this approach only requires information that would typically be available in applications, i.e.~locations of candidate positions and signal strength measurements between pairs of nodes. 

Consider a network of $M$ wireless nodes labeled $n_1, n_2, \ldots, n_M$, and $M$ candidate positions $p_1, p_2, \ldots, p_M$. It is unknown at which position each node is located. Nodes are equipped with radio transceivers and can exchange messages among them and thus RSSI between pairs of nodes can be obtained and sent to a backhaul server for post-processing. For localization purposes the signal strength between nodes $n_i$ and $n_j$ is converted into entry $D_{ij}$ of the distance matrix using a suitable propagation model. 

We now construct a similarity matrix $\bf C$ as
\begin{align}
C_{ij}=
\begin{cases}
      k(D_{ij}) & i \neq j \\
      0 & i = j
\end{cases} 
\end{align}
where $k$ is an appropriate kernel. Here we use the Gaussian kernel
\begin{equation}
    k(d)= \exp{(-\frac{d^2}{\sigma^2})}
\end{equation}
with
\begin{equation}
    \sigma^2= \frac{1}{M^2}\sum_{i,j}{D_{ij}}^2
\end{equation}
It is important that a short ranged kernel is used as the most salient information will be contained in the estimated distances between close nodes. 

We now regard the similarity matrix as the adjacency matrix of a weighted network and construct the network's row-normalized Laplacian
\begin{align}
L_{ij}=
\begin{cases}
      -\frac{C_{ij}}{\sum_{k}C_{kj}} & i \neq j \\
      1 & i = j
\end{cases} 
\end{align}
This matrix is related to the transition matrix of a random walk in a network, and describes diffusion processes on the network nodes \cite{coifman2005geometric}.

To construct a natural coordinate system for the nodes we consider spectral properties of $\bf L$. It is guaranteed that $\bf L$ has real non-negative eigenvalues. The number of zero eigenvalues is identical to the number of components in the network \cite{chung1997spectral} and hence we expect to find exactly one zero eigenvalue. The eigenvector corresponding to this eigenvalue does not carry any information and is ignored in the following. 

The most relevant eigenvalues for our purpose are the smallest non-zero eigenvalues as they carry information about the major dimensions of the system. We define $i$th eigenvector to be the eigenvector with the $i$th smallest non-zero eigenvalue. 

We interpret the entries of eigenvectors of $\bf L$ as coordinates along new coordinate axes \cite{coifman2006diffusion}. Note that the eigenvectors of the $M\times M$ matrix $\bf L$ have $M$ entries, i.e.~one entry per node. We store the coordinates in matrix $\bf N$ such that $N_{ij}$ is the $i$th entry of the $j$th eigenvector, corresponding to the $jth$ coordinate of node $n_i$. 

In typical use cases from applications the matching problem is effectively two dimensional such that the first two eigenvectors of the Laplacian already capture sufficient information. Consider for example a factory floor where the variation is much larger in the two horizontal dimensions than in the vertical. However, it is required to take the third eigenvector into account in problems where the third dimension is relevant.

In some other cases, taking other eigenvectors into account may also be necessary. The diffusion map is essentially a harmonic analysis of networks and its use in localization relies on using eigenvectors that span the main dimensions of physical space. The first eigenvector always spans the longest dimension of the system. However, if the positions are in a layout that has significantly different length scales (for instance, if the positions are on the boundary of a narrow rectangle), the eigenvector that encodes information about the location in the secondary direction may not be the second eigenvector as harmonics of the major direction might have smaller eigenvalues. In such cases taking a small number of additional eigenvectors beyond the second, including the eigenvector for the shorter dimension, into account solves the problem. Our results (below) indicate that this is a very minor issue in practice as even in very long and thin geometries accurate matching is possible without the eigenvector for the short dimension unless the nodes form a perfect lattice. In the examples below we use the two eigenvectors of $\bf L$ corresponding to the two dimensions of the layout such that $\bf N$ is an $M\times 2$ matrix, unless noted otherwise.

We repeat the above process for the predefined positions. The distance matrix is now formed using the distances between the known positions
\begin{align}
    D'_{ij}= \norm{p_i-p_j}.
\end{align}
We construct the corresponding similarity matrix and normalized Laplacian as above and use the same Laplacian eigenvectors as our new coordinates for the positions. We denote the $j$th coordinate of position $p_i$ as $P_{ij}$ which is analogous to $N_{ij}$ for the nodes. 

Using the diffusion map we have found the new coordinates of the nodes $\bf N$ and positions $\bf P$. The matching can now be done in terms of these new coordinates. Because the identification of the new coordinates is based on eigenvector computation, information about the sign of the axes is lost. Hence, coordinates for nodes can have opposite signs (but are not scaled) with respect to the corresponding coordinates for positions. While this slightly complicates matters, it is a very small concern for the actual application. If we know the position of one of the nodes, we can compare the signs of the node's coordinates with those of the corresponding position. If any of the coordinates differs in sign, we invert the signs of the respective coordinate entries for all nodes (i.e.~the entries in the respective column of $\bf N$). 

Even in the case where we don't have one known node we can run the matching multiple times for the different axes coordinate signs and compare the quality of matching. If we are working in two dimensions that means the matching needs to be done 4 times (axis 1 inverted, axis 2 inverted, both inverted, none inverted) and the orientation that leads to the best match (see below) is picked as the result. In practice, this solves the problem unless the configuration of positions is fundamentally ambiguous (e.g.~forming a symmetric lattice), in which case the correct matching can be picked from the, typically 4, alternatives by testing the assignments in one node. 

For the matching we compute the Euclidean distances $E_{ij}$ between the locations of node $n_i$ and position $p_j$ in terms of the diffusion coordinates. We then organize nodes and positions into assignment pairs such that the sum of the distances within pairs is minimized. For this purpose we use the Kuhn-Munkres algorithm, commonly referred to as the Hungarian algorithm \cite{kuhn1955hungarian,munkres1957algorithms}.

The Hungarian algorithm is used to find a minimum weight matching in the bipartite graph where nodes and positions are vertices and $E_{ij}$ is the weight of the edge that can be placed between $n_i$ and $p_j$. The algorithm then uses dynamic rebalancing of the edge weights to identify a set of links of minimal weight that connects each node to exactly one position.  

For the method proposed here, the Hungarian algorithm is the rate-limiting step that determines computational complexity. The algorithm can be implemented with a time complexity of $\mathcal{O}(M^3)$. For very large problems it might be advantageous to switch to alternatives to the algorithm such as one proposed by Ramshaw and Tarjan \cite{ramshaw2012minimum} with a running time of $\mathcal{O}(M^2\sqrt{M}log{}(M))$. However, we expect that the efficiency of the method, as proposed here, will be sufficient for most applications. A reasonably efficient implementation on a standard desktop computer should be able to quickly match many thousands of nodes. 

\section{Simulation Results}
As a first test we consider a system with 58 positions arranged in a plausible layout for a factory floor (Fig.~1a). Noiseless RSSI values are calculated from the distance matrix between the nodes. Gaussian noise is then added to RSSI values and the noisy distance matrix $\bf D$ is calculated. The SNR is calculated as signal mean over the standard deviation of the noise. Localization accuracy is evaluated as proportion of nodes assigned to correct positions. As expected, using one node to align the eigenvectors reduced the computation time but lead to exactly the same results as picking the best of the four candidate assignments. The algorithm yields good accuracy even at moderate SNRs (Fig.~1).
At an SNR greater than 5, it is almost certain that the correct matching is retrieved.
\begin{figure} 
\centering
\begin{subfigure}{.5\textwidth}
  \centering
   \includegraphics[scale = 0.22]{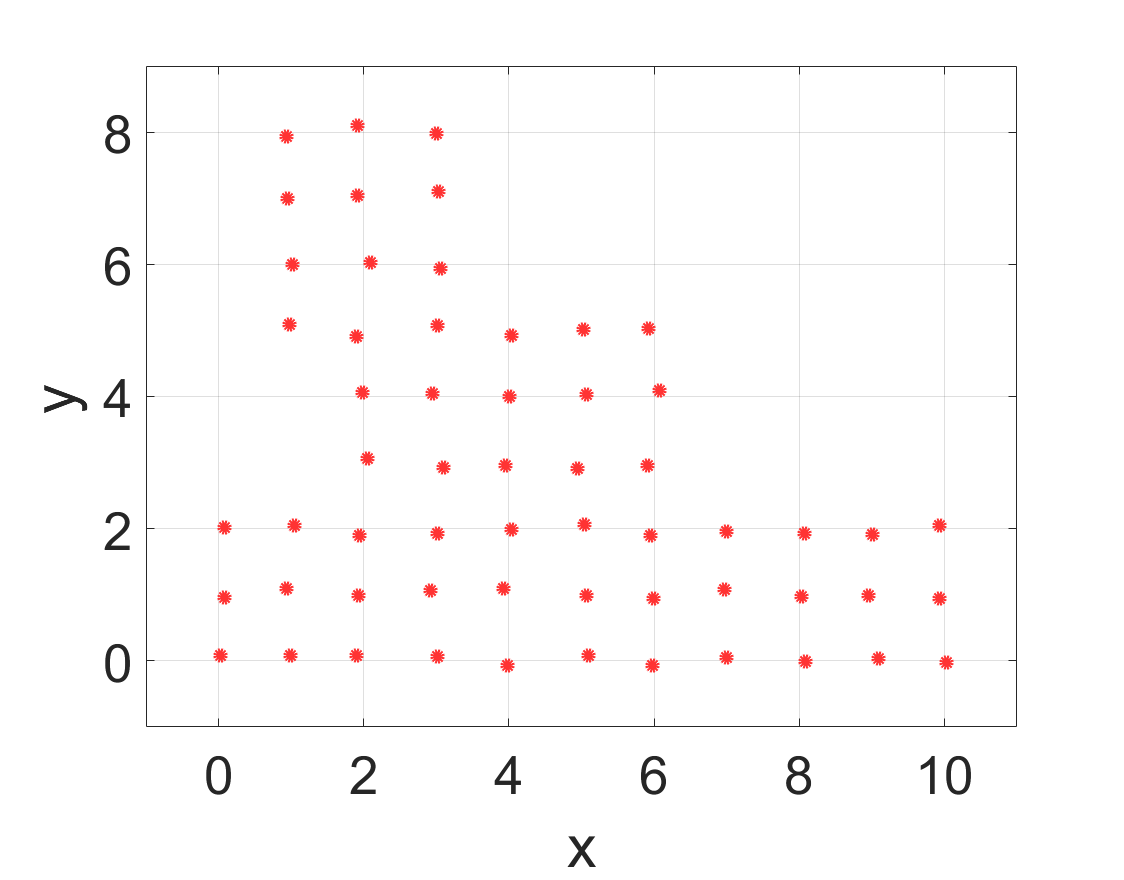}
    \caption{}
\end{subfigure}%
\begin{subfigure}{.5\textwidth}
  \centering
  \includegraphics[scale = 0.22]{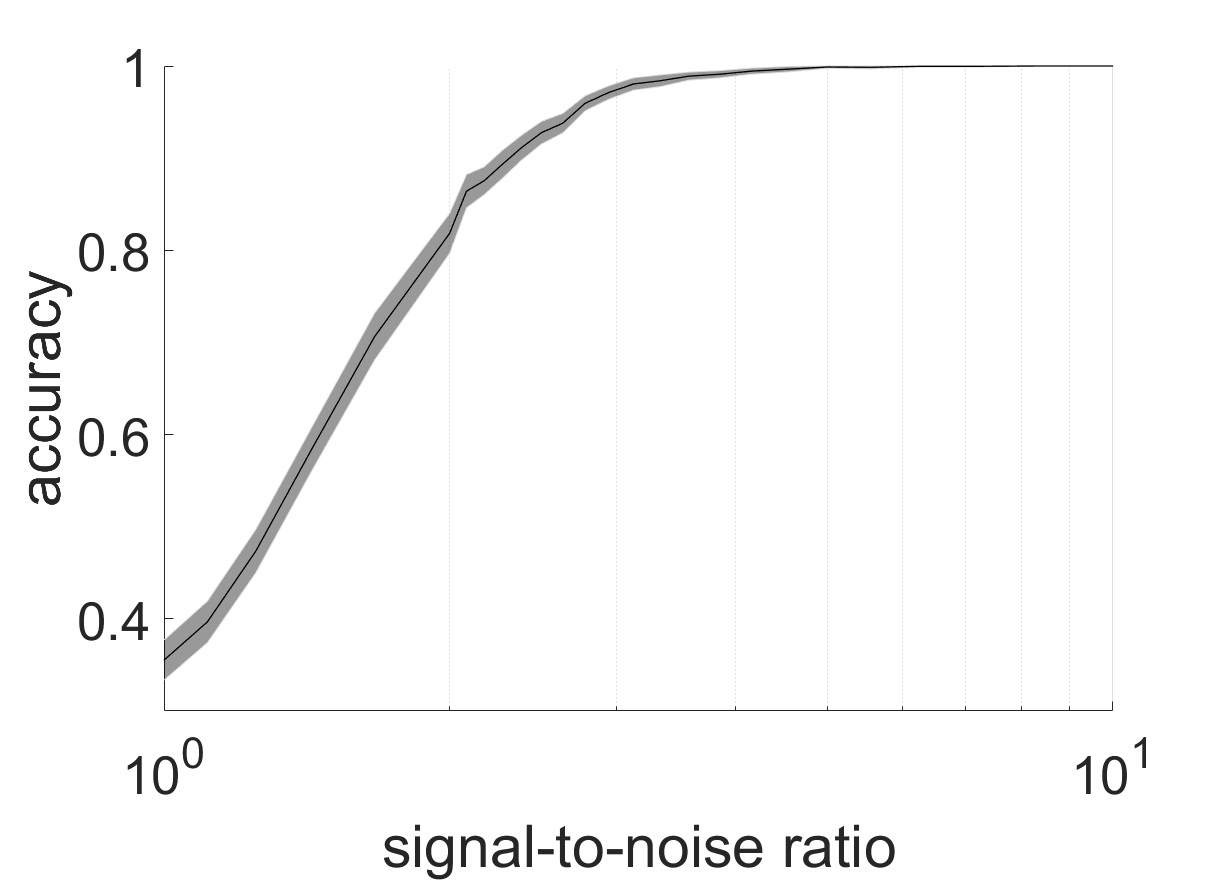}
    \caption{}
\end{subfigure}
\caption{Performance in a realistic scenario. Shown are  node positions (a) and the achieved accuracy in terms of signal-to-noise-ratio (b). The curve is a mean over 100 realizations of the noise. The area around the curve represents a 99\% confidence interval. In such realistic scenarios the method achieves perfect or near perfect matching results even at low signal-to-noise ratios of around 5.}
\end{figure}

To understand possible sources of failures we now investigate some intentionally difficult cases. We consider 4 different layouts where 80 positions are placed in a 2-dimensional grid (grid layout) and randomly in a two dimensional plane (random 2D layout), and 81 positions are placed equidistantly along both coordinate axes (uniform biaxial layout) and randomly along the coordinate axes (random biaxial layout).  The two random layouts are harder to match because they contain some positions that are very close together and hence easy to confuse. Proximity of nodes also explains why accuracy is less in the biaxial layouts. However, in all of these cases the method still achieves perfect or near-perfect results at moderate SNRs.  
\begin{figure} 
\centering
\begin{subfigure}{.25\textwidth}
  \centering
   \includegraphics[scale = 0.13]{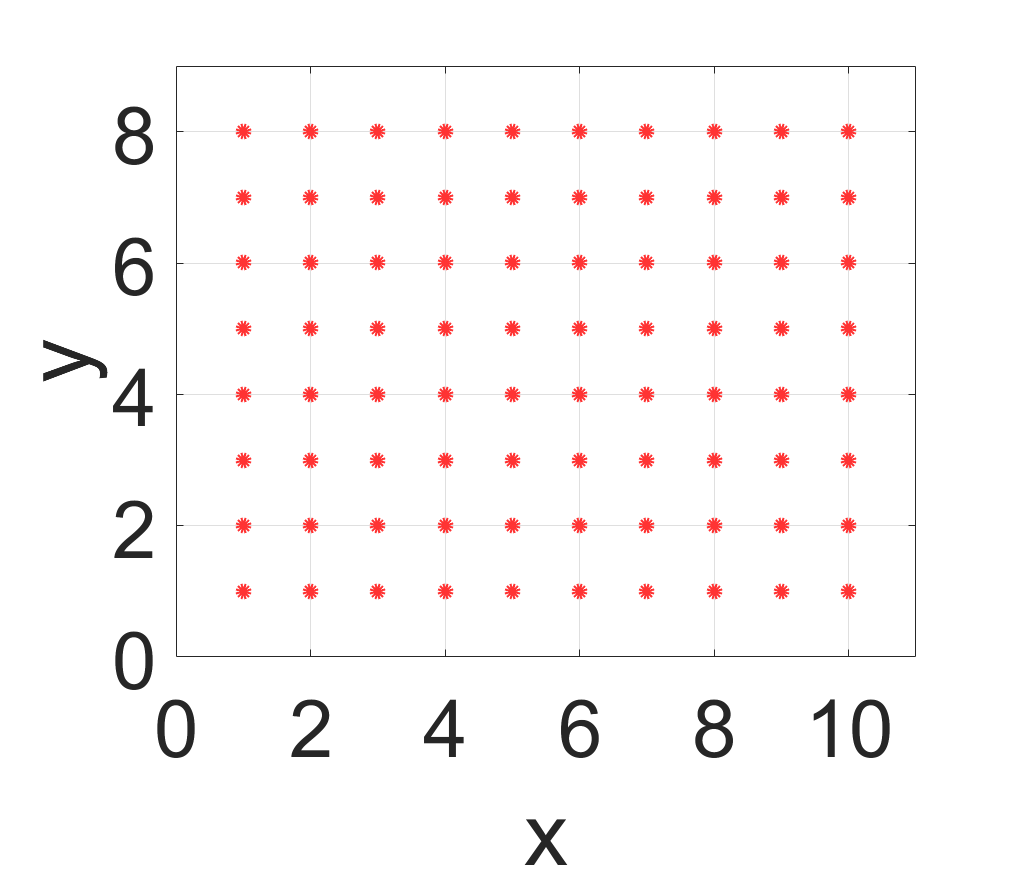}
    \caption{}
\end{subfigure}%
\begin{subfigure}{.25\textwidth}
  \centering
  \includegraphics[scale = 0.17]{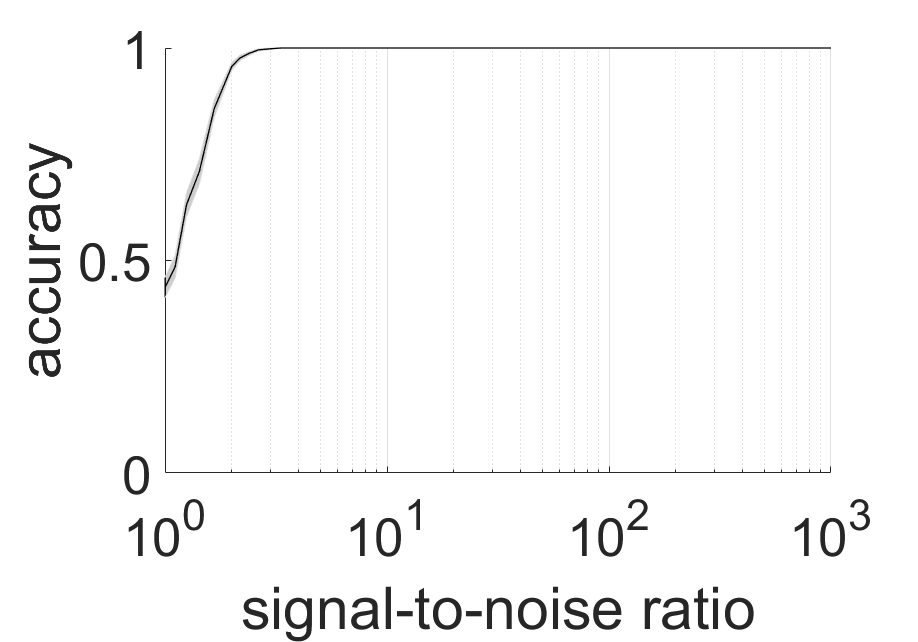}
    \caption{}
\end{subfigure}%
\begin{subfigure}{.25\textwidth}
  \centering
  \includegraphics[scale = 0.13]{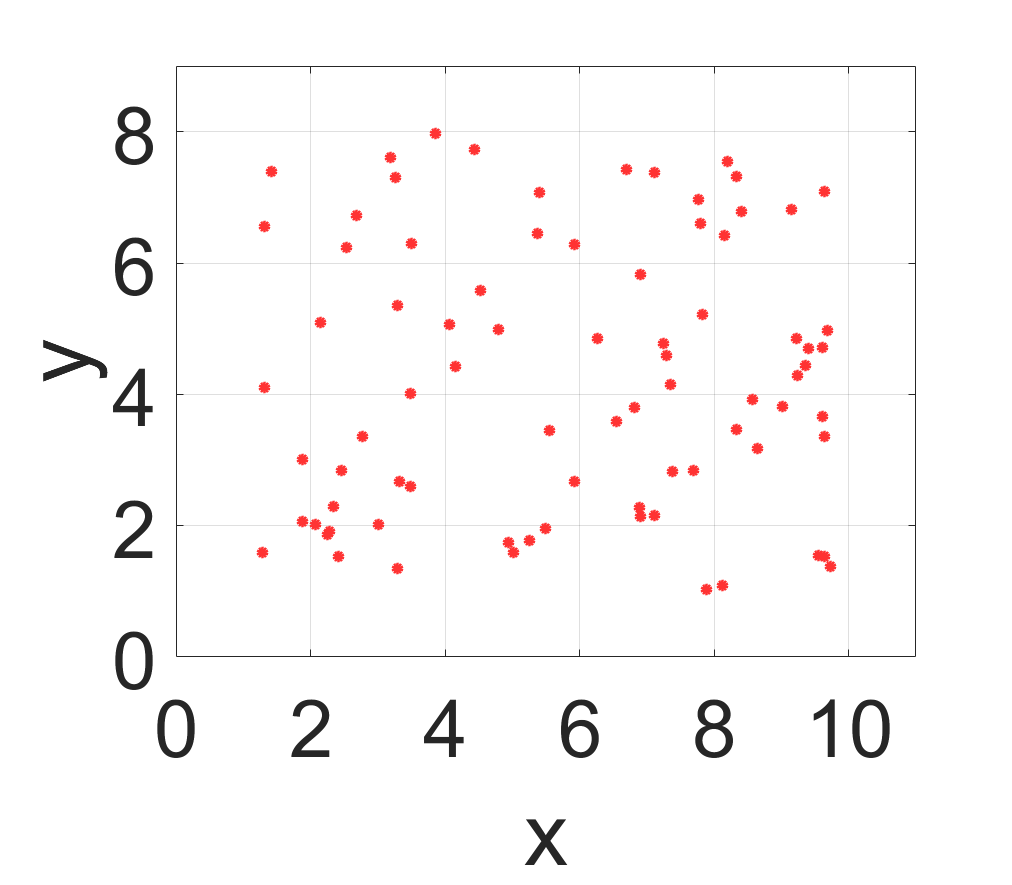}
    \caption{}
\end{subfigure}%
\begin{subfigure}{.25\textwidth}
  \centering
  \includegraphics[scale = 0.17]{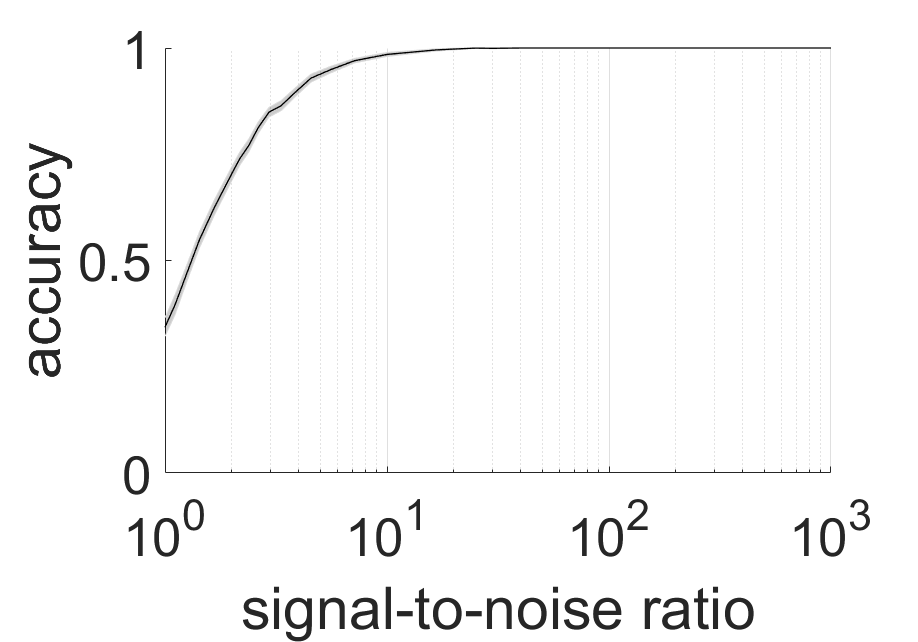}
    \caption{}
\end{subfigure}
\begin{subfigure}{.25\textwidth}
  \centering
   \includegraphics[scale = 0.13]{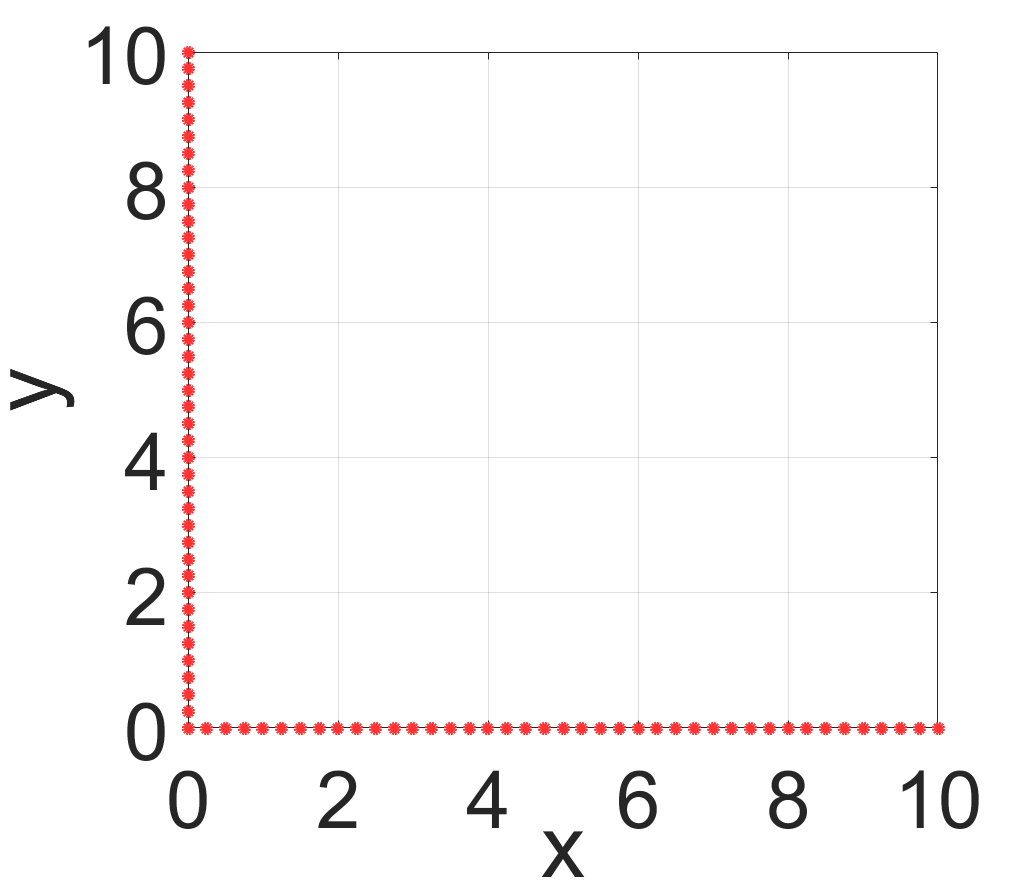}
    \caption{}
\end{subfigure}%
\begin{subfigure}{.25\textwidth}
  \centering
  \includegraphics[scale = 0.17]{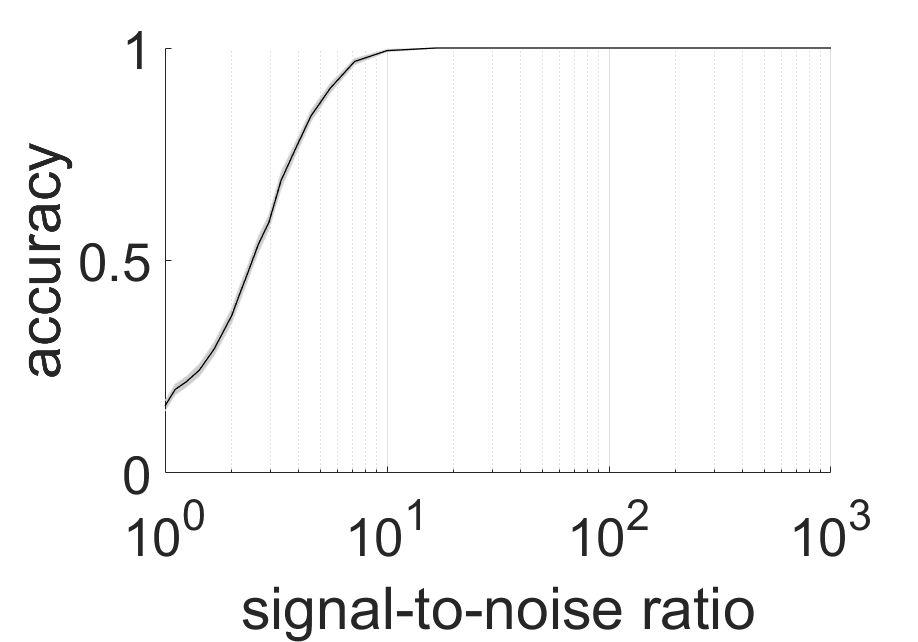}
    \caption{}
\end{subfigure}%
\begin{subfigure}{.25\textwidth}
  \centering
  \includegraphics[scale = 0.13]{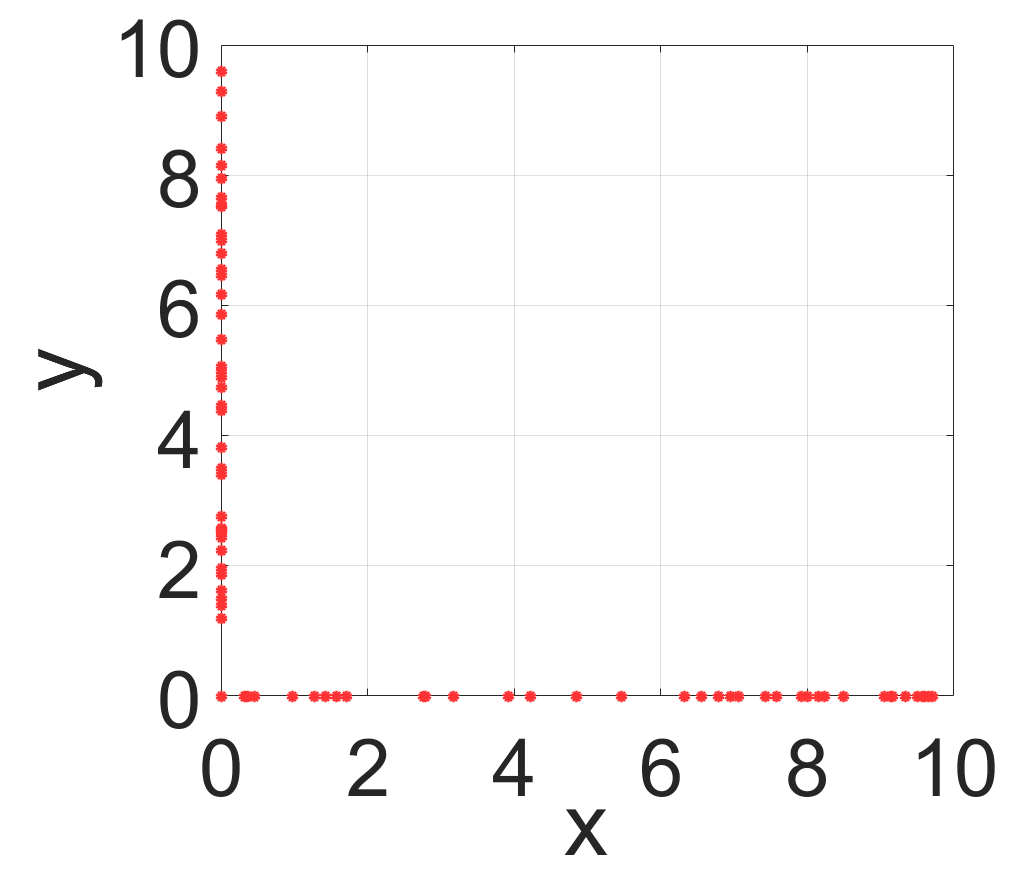}
    \caption{}
\end{subfigure}%
\begin{subfigure}{.25\textwidth}
  \centering
  \includegraphics[scale = 0.17]{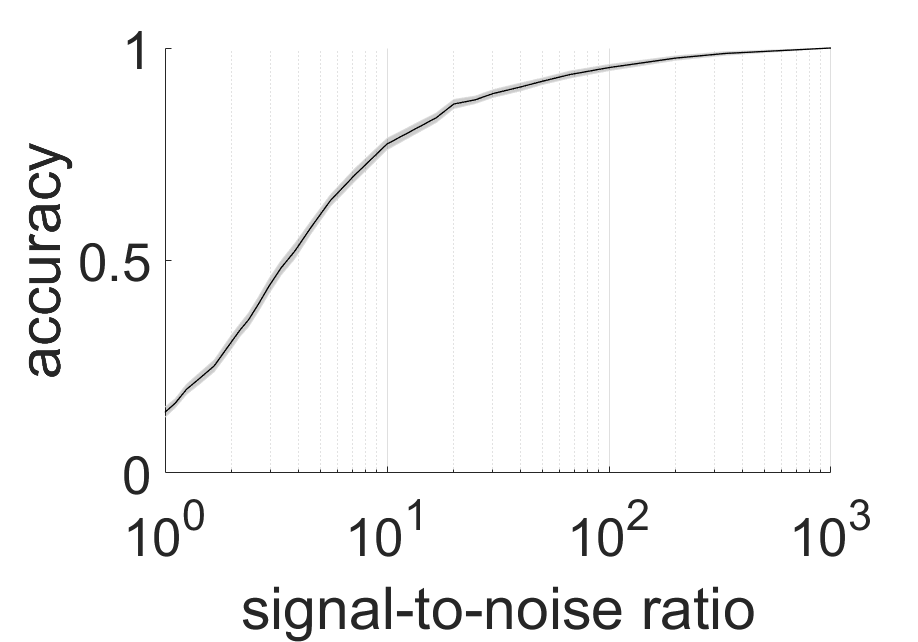}
    \caption{}
\end{subfigure}
\caption{Performance in different artificial scenarios. Shown are node positions and matching accuracy analogously to Fig.~1. Matching is harder in biaxial layouts (e-h) than in uniform layouts. Highly random layouts (c,d,g,h) introduce an additional difficulty because they contain some positions that are very close together. Even in these intentionally difficult scenarios good matching results can be achieved if the signal-to-noise ratio is sufficiently high.   
}
\end{figure}

In Sec.~2 we mentioned that complications can arise from the presence of harmonic eigenvectors. To illustrate these we study a system of 40 positions in a strip layout (Fig.~3a). In this case the first eigenvector contains information about the position of nodes along the primary axis of the strip. Hence this eigenvector cannot be used to differentiate between nodes that only differ in the secondary axis position (Fig.~3b).   

The second and third eigenvectors are higher harmonics of this eigenvector and thus contain essentially the same information. While these eigenvectors can be taken into account for more accurately locating a node in the primary direction, even using all 3 eigenvectors does not let us distinguish between nodes that lie side-by-side in the strip. Hence the matching accuracy plateaus at around 50\% if these eigenvectors are used in the matching.  

The first eigenvector that contains information about the position along the secondary axis is eigenvector 4. Hence using either the first and fourth eigenvector, or all of the first 4 eigenvectors yields good matching results.  

We note that such problems due to harmonics only arise in the relatively artificial situation where the two rows of the strip are perfectly side-by-side. As we shift one of the rows even by small amounts, using only the first eigenvector yields an increasingly accurate matching, possibly even surpassing the accuracy for when considering the first and the fourth eigenvectors.

We expect that the problems caused by the presence of harmonics will only play a minor role in applications. However, the possibility that such problems might occur highlights the need to select appropriate eigenvectors for the matching. The choice of eigenvectors to use can be made by checking whether a given set of eigenvectors can resolve all positions. Thus the backhaul server could determine a suitable set of eigenvectors once the blueprint is made available. 
\begin{figure} 
\centering
\begin{subfigure}{.33\textwidth}
  \centering
   \includegraphics[scale = 0.21]{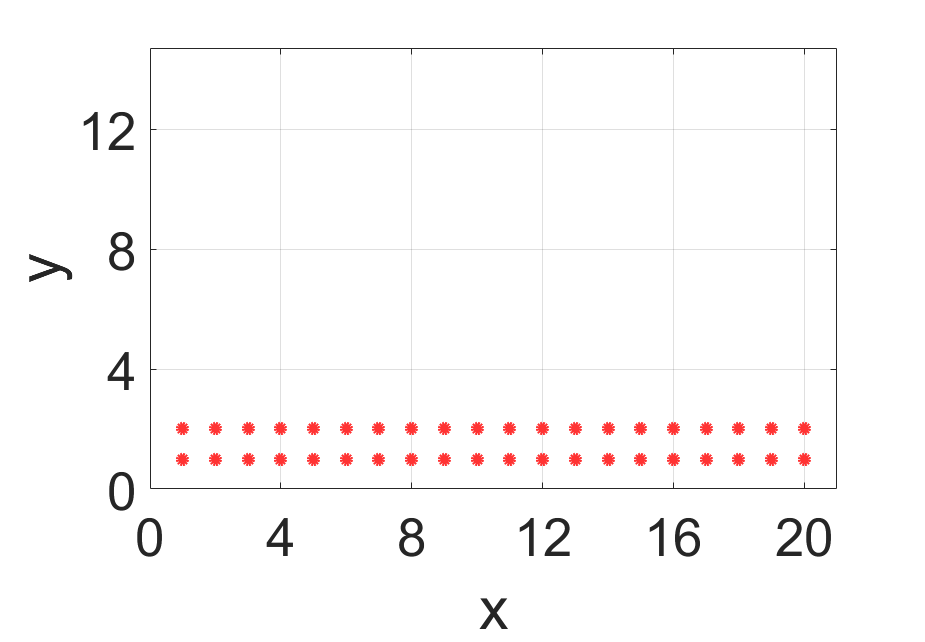}
    \caption{}
\end{subfigure}%
\begin{subfigure}{.33\textwidth}
  \centering
  \includegraphics[scale = 0.21]{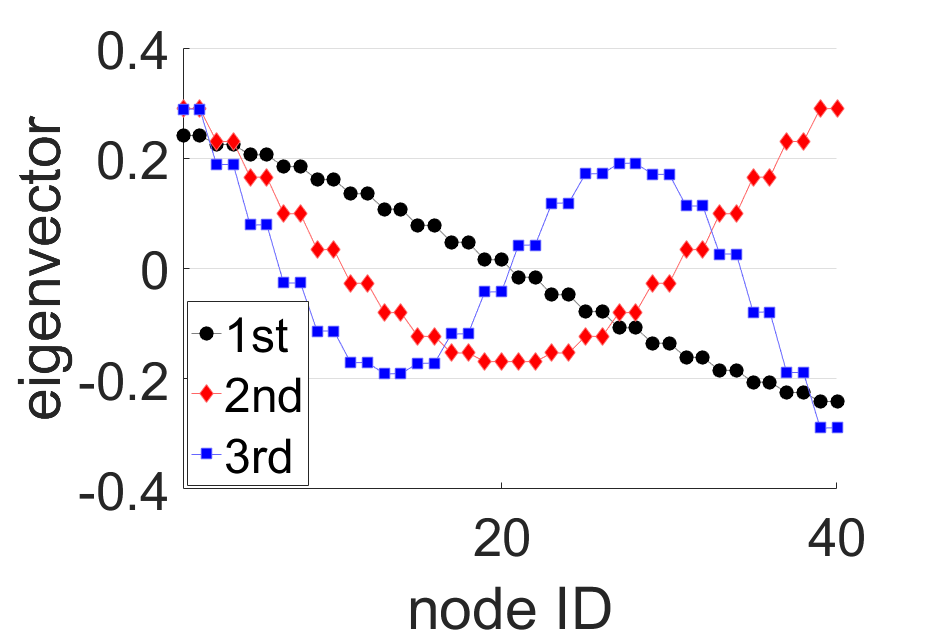}
    \caption{}
\end{subfigure}%
\begin{subfigure}{.33\textwidth}
  \centering
  \includegraphics[scale = 0.21]{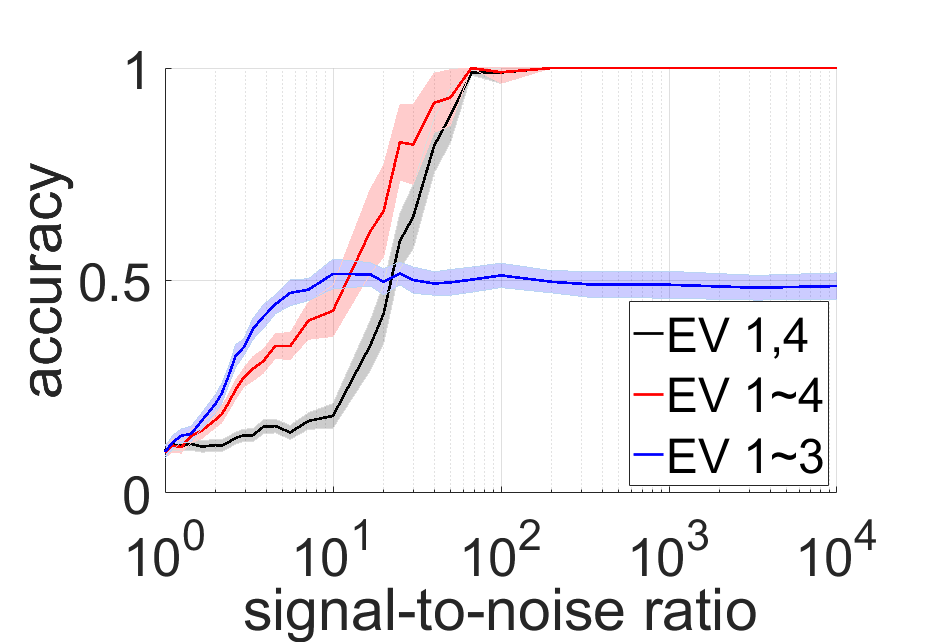}
    \caption{}
\end{subfigure}%
\caption{Usage of additional eigenvectors. If positions form long and thin lattices (a) information about the location along the shorter lattice direction is only contained in higher eigenvectors (b), as the first three eigenvectors assign almost the same value to pairs of points. Hence the accuracy plateaus at around 50\% if only the first three eigenvectors (blue) are used for matching (c). However, using the first and the fourth eigenvectors (black) or all of the first 4 eigenvectors (red) yields good results.}
\end{figure}

\begin{figure} 
\centering
\begin{subfigure}{.5\textwidth}
  \centering
   \includegraphics[scale = 0.3]{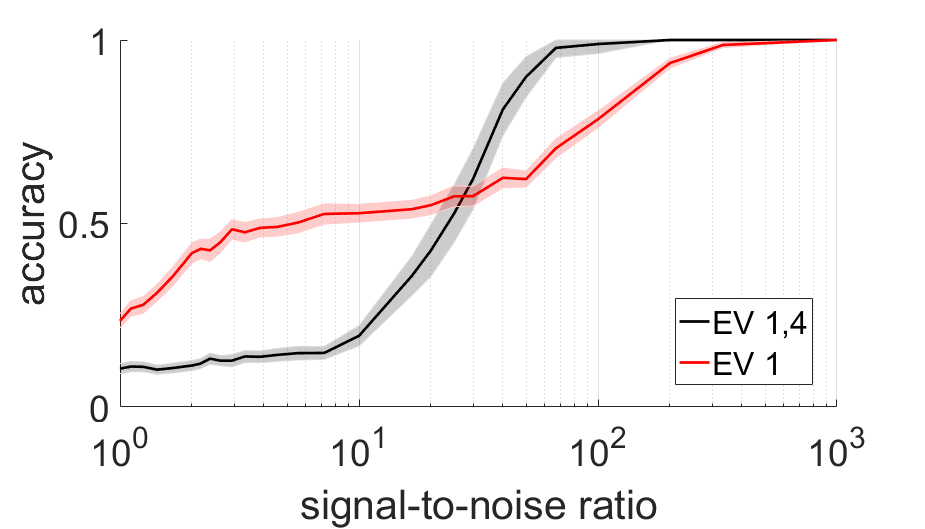}
    \caption{}
\end{subfigure}%
\begin{subfigure}{.5\textwidth}
  \centering
  \includegraphics[scale = 0.3]{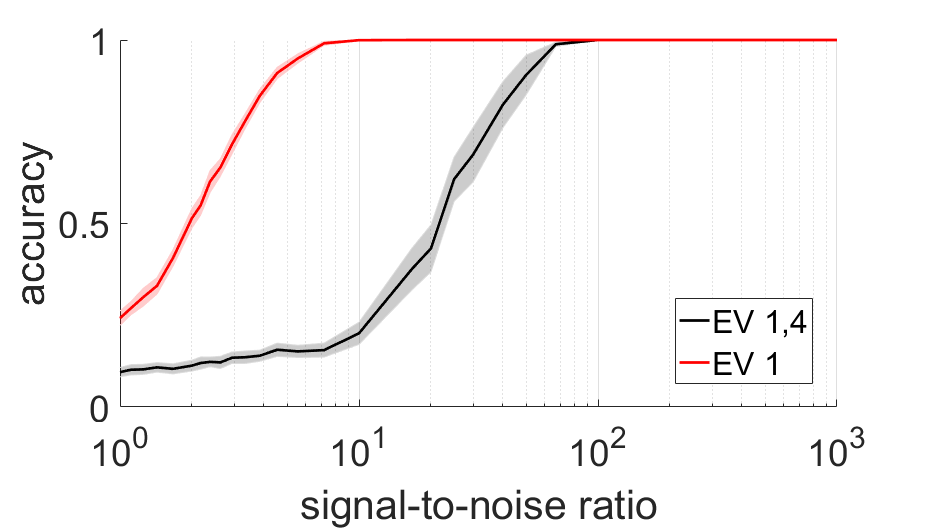}
    \caption{}
\end{subfigure}
\caption{Small shifts in a lattice mitigate the effect of harmonics. We consider the layout from Fig.~3a, but shift one of the rows of positions by either 0.01 units (a) or 0.5 units (b). While this has little effect on the matching using eigenvectors 1 and 4 (black) already a small shift significantly improves the matching when only eigenvector 1 is used (red). At small signal-to-noise ratios or for larger shifts, using the first eigenvector yields better results.}
\end{figure}

In our final simulations we investigate problems in 3D. We consider 120 positions that are either placed randomly or form a 3-dimensional grid. Location matching, using the first 3 eigenvectors, yields in both cases accurate results even at low SNR. The accuracy thus surpasses that of the 2D layouts. This increase in accuracy is explained because the network of measurements is denser in 3D (i.e.~lower effective diameter of the node graph) and a lesser chance that nodes are placed very closely together in the random layout.   
\begin{figure} 
\centering
\begin{subfigure}{.5\textwidth}
  \centering
   \includegraphics[scale = 0.3]{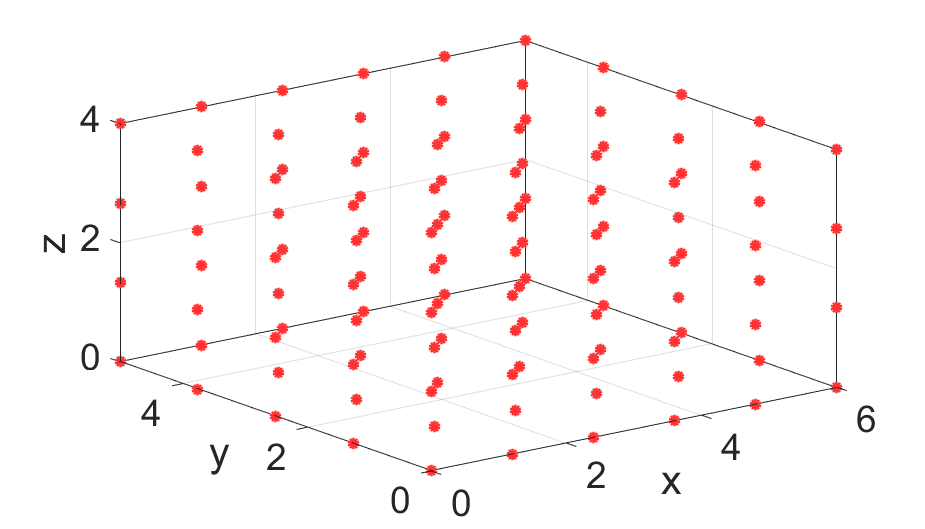}
    \caption{}
\end{subfigure}%
\begin{subfigure}{.5\textwidth}
  \centering
  \includegraphics[scale = 0.3]{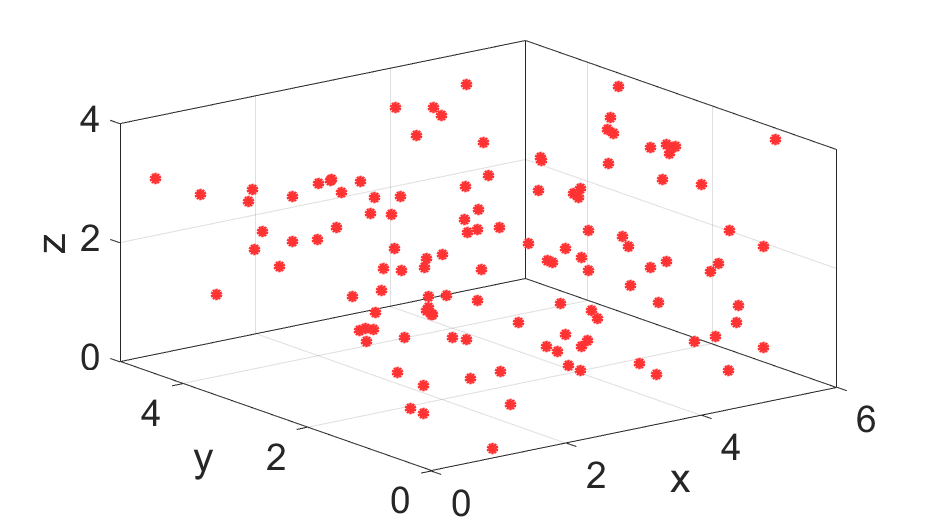}
    \caption{}
\end{subfigure}
\begin{subfigure}{1\textwidth}
  \centering
  \includegraphics[scale = 0.3]{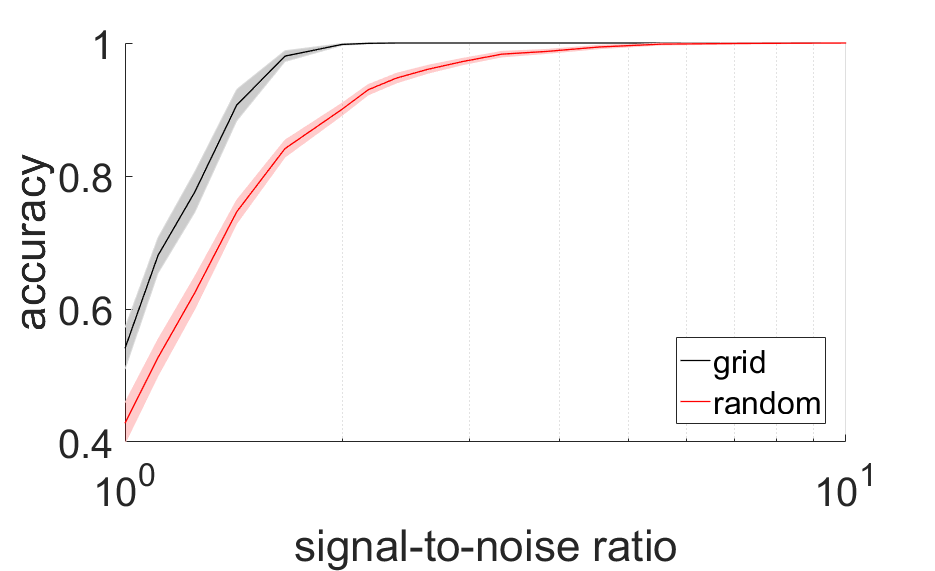}
    \caption{}
\end{subfigure}
\caption{Localization performance in 3D problems. Node positions for grid (a) and random (b) layouts are shown. The accuracy plots (c) illustrate that the method can be implemented in solving 3D problems.}
\end{figure}

\section{Conclusion}
In this paper we proposed a solution to the Wireless Localization Matching Problem. 
The key innovation was to use diffusion maps to embed measurements 
in a low dimensional manifold spanned by natural coordinate axes of the problem. 
We demonstrated in numerical simulations that the proposed algorithm can solve the matching problem without errors in realistic examples if the signal-to-noise ratio exceeds a threshold that is typically between $10^1$ and $10^2$. We believe this is sufficient to allow robust matching in applications. 

We emphasize that our algorithm does not require information beyond the blueprint of potential positions of equipment and pairwise signal strength measurements between proximal wireless sensors. In contrast to previous approaches \cite{keller2011diffusion} for related problems, the method proposed here does not require anchor nodes, except in the case of fundamentally ambiguous layouts, where one node needs to determine the orientation of eigenvectors. It thus uses only information that would typically be accessible in the envisioned applications. Likewise, the numerical demand is such that even for very large systems it can be met with readily available desktop hardware. 

When applying the algorithm, care has to be taken to take the right number of eigenvectors into account. However, it is easy to work out the right number by mapping the positions and testing for modal relationships among the first few eigenvectors. While it is thus essential that applications of the method should include a preprocesing step in which suitable eigenvectors are picked, this step could be straightforwardly integrated in software. 

The examples considered in our tests focused on intentionally difficult cases. Configurations of positions in the real world are likely to result in more accurate assignments than some of the extreme scenarios considered here. We expect that similar to the example in Fig.~1 perfect matching can already be achieved at a signal-to-noise ratio of about 5.

We hope that the proposed method will help to realize the future applications of wireless localization. We anticipate that future refinements of this approach will lead to additional improvements. Such refinements may include the use of improved models in the construction of the distance matrix or the use of a thresholding step, which is commonly used in other applications of the diffusion map. However, the fine-tuning of the additional parameters introduced by these refinements as well as the benefit conveyed by them will likely depend on the specific application.    
\section*{Acknowledgements}
OG would like to acknowledge funding from the EUs H2020 research and innovation programme under the Marie Skłodowska-Curie grant agreement NEWSENs, No 787180.

\section*{References}

\bibliographystyle{unsrtnat}
\renewcommand{\bibsection}{}
\bibliography{references}

\end{document}